\documentclass[final,3p,times,onecolumn]{elsarticle}
\usepackage{amsmath,amssymb,amsfonts,scalerel}
\usepackage{blindtext}
\usepackage{graphicx}
\usepackage{adjustbox}
\usepackage{epsfig}
\usepackage{float}
\usepackage{subfig}
\usepackage{xcolor}
\usepackage[colorlinks=true,citecolor=blue,linkcolor=blue,urlcolor=blue]{hyperref}
\usepackage{setspace}
\usepackage{multirow}
\usepackage{multicol}
\usepackage{url}
\usepackage{hyperref}
\usepackage{rotating}
\usepackage{geometry}
\biboptions{sort,compress}
\journal{J.~Applied Physics}

\begin{document}

\begin{frontmatter}

\title{Study of novel properties of graphene-ZnO
       heterojunction interface using density functional theory}

\author{H.D.~Etea}
\author[]{K.N.~Nigussa\corref{cor1}}
\cortext[cor1]{Corresponding author:\ kenate.nemera@aau.edu.et\ 
               (K.N.~Nigussa)}

\address{Department of Physics,\ Addis Ababa University,\ P.O. Box 1176,\ 
         Addis Ababa,\ Ethiopia}
         
\begin{abstract}
Studies of the structural, electronic, and optical\
characteristics of the interfaces between graphene\ 
and ZnO polar surfaces is carried out using\ 
first-principles simulations.\ At the interface,\ 
a strong van der Waals\ force is present,\ and\ 
because of the different\ work functions of\ 
graphene and ZnO,\ charge transfer takes place.\ 
Graphene's superior\ conductivity is not impacted\ 
by its interaction\ with ZnO,\ since its Dirac\ 
point is unaffected despite\ its adsorption on ZnO.\ 
In hybrid systems,\ excited electrons with energies\
between 0\ and 3 eV (above Fermi energy)\ are primarily\ 
accumulated on graphene.\ The\ calculations offer a\ 
theoretical justification\ for the successful operation\ 
of graphene / ZnO\ hybrid materials as\ photocatalysts\ 
and solar cells.\ ZnO semiconductor is\ found to be\ 
a\ suitable material\ with modest band gap,\ ($\sim$ 3 eV),\ 
having high\ transparency in visible\ region and a\ 
high optical conductivity.   
\end{abstract}


\end{frontmatter}

\section{Introduction\label{sec:intro}}
The ultimate spintronics\ device can be thought\ 
of as\ the interface\ between materials since\ 
it allows for\ new design possibilities and\ 
physical features\ that are not possible in\ 
the individual bulk\ materials~\cite{MVDLH17}.\ 
Rashba-Edelstein\ spin-to-charge conversion\ 
and spin-momentum\ locking at the surface of\ 
topological insulators\ are two emerging\ 
interfacial phenomena\ caused by spin-orbit\ 
coupling\ ($\rm{SOC}$).\ Interfaces naturally\ 
break\ spatial inversion\ symmetry,\ which\ 
causes\ an electronic\ band Rashba\ $\rm{SO}$ splitting\ 
that is often\ higher than\ in bulk.\ In\ 
particular,\ $\rm{SO}$ effects\ at oxide\ interfaces\ 
are crucial\ for low-power\ spintronics\ 
applications due to\ the expected\ long\ 
carrier\ lifetime and\ high Rashba\ 
coefficient~\cite{JVBM18}.\ Carbon-based\ 
nanomaterials are\ prime candidates for\ 
spin-based\ devices,\ due to their long\ 
spin coherence\ length (up to~$10^{4}~ps$)\ 
and high Fermi\ velocity~\cite{SRoche2015}.\ 
Since the\ successful exploitation\ of\ 
single-layer\ graphene in\ 2004 by Geim\ 
and\ his co-workers~\cite{SGMDYVGF04},\ 
the investigations\ on the interaction\ 
of graphene and\ other nanomaterials\ 
have become\ a growing\ trend due to\ 
the extraordinary\ properties of\ 
graphene,\ such as the\ Dirac\ 
electrons near\ the K point,\ 
room-temperature\ quantum Hall\ effect,\ 
and high mobility\ of carrier\ electrons~\cite{MNKSEJG08}.\ 
Research on the\ interaction of graphene\
with BN~\cite{DSXRHJ11},\ SiC~\cite{MAP07},\ 
metals~\cite{NBFM2006},\ or metal oxides~\cite{HMZ2009}\ 
has been done\ both experimentally\ and\ 
computationally\ thus far.\ 
The study of\ graphene-based interfaces is\ 
of great importance\ in materials production~\cite{LiX2009},\ 
device fabrication~\cite{ACV10},\ and electrical\ 
measurement~\cite{HNSG08},\ because the contact\ 
of other materials\ with graphene may\ 
modify the\ physical and\ chemical properties\ 
of graphene and\ then influence\ its\ 
performance\ in devices.\\
In recent years,\ zinc oxide\ (ZnO) has\ 
attracted great\ attention in\ the field\ 
of nanodevices\ as an important\ semiconductor\ 
because of its\ unique optical,\ electronic,\ 
and magnetic\ properties,\ including large\ 
excitation binding\ energy (60~meV),\ wide\ 
band\ gap (3.4~eV),\ and unique piezoelectric\ 
properties~\cite{zALTRDVSM05}.\ A review of\ 
the literature\ reveals that the\ wide band\ 
gap,\ $>$~3~eV,\ ZnO semiconductor\ is a\ 
suitable\ material with\ great transparency\ 
in the\ visible region\ and good optical\ 
conductivity.\ 
Some of ZnO\ properties are\ improved when it\ 
is synthesized in\ the form of\ nanosized material\ 
and,\ for this reason,\ a wide range of ZnO\ 
nanostructures have\ been lately\ obtained~\cite{JNW06}\ 
such as\ nanohelixies,\ nanobows,\ nanorings,\ 
nanowires,\ and nanocages,\ which are promising\ 
candidates for\ gas sensors,\ solar cells,\ 
field effect transistors,\ photocatalysts,\ 
and so on~\cite{ZKYMNAT05}.\\
These materials\ display unique\ features such\ 
as greater photovoltaic\ properties than graphene\ 
or bulk ZnO alone in solar cells~\cite{YWXHQBZ10}\ 
and photocatalysts~\cite{KGLP12}.\ Despite the\ 
varied morphologies\ of graphene hybridizations\ 
with ZnO nanoparticles\ and vertical nanostructures,\
it is generally\ accepted that\ graphene's high\ 
specific surface\ area facilitates\ the loading\ 
of dyes\ in photocatalysis,\ and that charge\ 
transfer processes\ that involve charge-hole\ 
separation and\ transfer from\ graphene to ZnO\ 
are facilitated\ by graphene's\ higher conductivity.\ 
With significant\ experimental advancement,\ 
a computational investigation\ of graphene-ZnO\ 
hybrid systems has\ also been carried out~\cite{GZLY13}. 
In this work,\ the structural,\ interaction,\ and\ 
electronic properties\ of graphene-ZnO hybrid\ 
systems were\ investigated through density\ 
functional theory\ (DFT) computations\ to\ 
better understand\ the experimental results\ 
and hidden mechanisms.

The paper is organized as follows.~In\ 
the\ next\ section (sec.~\ref{sec:comp}),\ 
a\ detail account\ of the computational\ 
method\ is presented.\ Results\ and\ 
discussion are\ presented in\ 
section~\ref{sec:res},\ with\ the\ 
conclusion\ being presented in\ 
section~\ref{sec:conc}.
\section{Computational Methods\label{sec:comp}}

All calculations in this work were\ 
carried out using the GPAW\ code,\ where\ 
the detail of the capabilities\ is described\
in\ a literature~\cite{ERM10}.\ The exchange-correlation\ 
energy of the\ interacting electrons is\ 
expressed using\ the frozen-core full\ 
potential projector\ augmented wave (PAW)\ 
approach~\cite{PBJA94}\ and the Perdew-Burke-Ernzerhof\ 
(PBE) form~\cite{PBM96}.\ The electronic ground\ 
state was determined\ using the plane waves\ 
basis with a cut-off\ energy of 400 eV and\ 
the conjugate\ gradient algorithm,\ with the\ 
convergence threshold\ set at $5{\times}10^{-4}$ eV\ 
for energy\ and 0.01 eV/{\AA} for force.\ The\ 
Brillouin\ zone integrations\ were carried out\ 
using\ Monkhorst-Pack grids~\cite{MHP76}.\
We used k-point meshes of $6{\times}6{\times}6$\ 
for a wurtzite ZnO primitive cell,\ $6{\times}6{\times}1$\ 
for a graphene\ primitive cell,\ ZnO $(001)$\ 
polar surfaces,\ and\ the supercells of\ the\ 
interface.\ A dipole\ correction~\cite{MGPM95}\ 
was\ applied to\ make the computations\ converge\ 
more\ quickly\ and to eliminate the\ artificial\ 
electrostatic\ field between\ periodic supercells.\

The\ electron\ wave-function\ is approximated\ 
using the\ implementation\ of the\ projector\ 
augmented\ wave~(PAW)\ method and\ as\ expanded\ 
in\ a\ plane wave\ basis\ set~\cite{Blochl94}\ 
having a\ band,\ a k-points,\ and a reciprocal\ 
lattice vector\ grid indices.\ The electron\ 
energy\ eigenvalues\ are\ obtained\ by\ applying\ 
Schr$\rm\ddot{o}$dinger\ equation and\ solving\ 
self-consistently\ via\ the\ Kohn-Sham\ 
scheme~\cite{KS65}.\   
The\ interactions\ of the valence\ electrons\ 
with\ the\ core electrons\ and nuclei\ is\ 
treated\ within\ a projector\ augmented wave\ 
(paw) data\ sets~\cite{Enkovaaraetal2010,MHJ2005}.\
The number of\ valence electrons\ considered\ 
for\ each\ element within\ the paw\ data sets\ 
is\ Zn~(3d$\rm^{10}$4s$\rm^{2}$),\  
O~(2s$\rm^{2}$2p$\rm^{4}$),\
C~(2s$\rm^{2}$2p$\rm^{4}$).\
Geometry\ relaxations\ are carried out\ 
using\ BFGS\ minimizer~\cite{BS82},\ where\ 
optimization\ of the\ atomic forces\ and\ 
the unit cell\ stresses\ is\ done within\ 
the concepts\ of the\ Hellmann-Feynman\ forces\ 
and\ stresses~\cite{RF39,NM85}\ calculated\ 
on\ the Born-Oppenheimer~(BO)~surface~\cite{WM91}.\  
The\ exchange-correlation\ energies are\ approximated\ 
within the\ generalized\ gradient approximation\ 
of PBE~\cite{PBE96}.\ The\ strongly\ 
correlated\ nature\ of\ $d$\ electrons\ 
of\ Zn\ were\ treated\ using\ Hubbard-like\ 
model~which is\ introduced into the Gpaw\ 
code according to~\cite{LAZ95,DBSHS98},\ 
where\ U-J=9.5~eV.\ 

Spin polarized\ calculation is\ applied\ 
where\ a\ magnetic moment\ on each atom\ 
is allowed to\ relax\ to\ optimum value.\ 
Density of states~(DOS)\ 
is calculated\ as a population\ of states\ 
available\ for occupation\ at a given energy.\ 

Equilibrium volumes and bulk\ modulus\ 
are\ calculated using Murnaghan\ 
equation of state~\cite{FDM44}.\
Cohesive energy\ is calculated as the\ 
energy required\ to break atoms of\ 
the solid into\ a constituent isolated\
atoms,\ as described elsewhere~\cite{SK2022}.\ 
In the models\ of the interface,\ 
vdW interactions\ are expected\ to be significant\ 
and thus the\ DFT-D2 method of\ Grimme~\cite{GS2006},\ 
which is successful\ for graphene/SiC\ interface\ 
structures~\cite{JTKN11} was\ adopted in\ 
this work.\ The total energy\ $(E_{total})$\ 
is represented as:\
\begin{equation}
E_{total}=E_{KS-DFT}+E_{vdW}
\end{equation} 
where\ $E_{KS-DFT}$ is\ the conventional\ 
Kohn-Sham\ DFT energy\ and\ $E_{vdW}$\ is the\ 
dispersion correction.\ Note that\ Vanin\ 
et al~\cite{VMK10}\ reported that\ the vdW-DF\ 
method\ proposed by\ Dion et al~\cite{DRH04}\ 
failed to\ reproduce the\ experimental observations\ 
of the\ metal-graphene interface.\ When doing a\ 
spin polarized\ computation,\ a magnetic moment\ 
on each atom\ is allowed to\ relax to its optimum\ 
value.\\
In this work,\ the ZnO wurtzite\ structure with\ 
unit cells of\ 6 atoms in\ $P6_{3mc}$ is taken\ 
into consideration.\ After these\ frameworks were\ 
set up,\ literature resources~\cite{RSJ2008},\ 
were carefully followed.\ The number of states\ 
that are\ occupied at a\ given level of energy\ 
is used to\ compute the density\ of states (DOS).\ 
Tetrahedron approach\ was used to do\ Brillouin\ 
zone integration,\ and it has been\ shown to be\ 
effective,\ particularly for\ calculations of\ 
excited states\ and dielectric functions~\cite{physb2022}.\ 

The\ optical\ response\ property~is\ analysed\ 
from dielectric\ function,\ which\ is\ 
given as\
\begin{equation}{\label{eq1}}
\varepsilon(\omega)=\varepsilon\rm_{1}
(\omega)+i\varepsilon\rm_{2}(\omega)
\end{equation}
The imaginary part\ $\varepsilon\rm_{2}(\omega)$\ 
is\ calculated\ from the density\ matrix\ 
of\ the\ electronic\ structure~\cite{HL87}\ 
according\ to\ the\ implementations\ by\ 
the\ group\ of\ G.\ Kresse~\cite{GHKFB2006,SK2006},\ 
$\&$\ given by\
\begin{equation}{\label{eq2}}
\varepsilon\rm_{2}(\omega)=\frac{8{\pi}^{2}e^{2}{\hbar}^{2}}
{\Omega {{\omega}^{2}}{m_{e}^{2}}} 
{\sum\limits_{k,v,c}}{w\rm_{k}}{{\mid}
\langle{\psi\rm_{k}^{c}}{\mid}{\bf u}{\cdot}
{\bf r}{\mid}{\psi\rm_{k}^{v}}\rangle{\mid}}^{2}
\delta(E\rm_{k}^{c}-E\rm_{k}^{v}-\hbar \omega), 
\end{equation}
where $e$ is the\ electronic charge,\ and\ 
$\psi\rm_{k}^{c}$\ and\ $\psi\rm_{k}^{v}$\ 
are\ the conduction band\ (CB)\ and\ valence\ 
band\ (VB)\ wave functions at k,\ respectively,\ 
$\hbar \omega$\ is the\ energy of the\ incident\ 
photon,\ ${\bf u}{\cdot}{\bf r}$\ is\ the\ 
momentum operator,\ $w\rm_{k}$\ is a joint\ 
density of states,\ $\&$\ $\Omega$\ is\ volume\ 
of\ the\ primitive cell.\

The\ Real\ part\ $\varepsilon\rm_{1}(\omega)$\ 
of\ the dielectric\ function\ can be\ found\ 
from\ the\ Kramer-Kronig\ equation~\cite{FW72}.\
\begin{equation}{\label{eq3}}
\varepsilon\rm_{1}(\omega)=1+\frac{2}{\pi}
P\int_{0}^{\infty}\frac{\omega^{'}
\varepsilon\rm_{2}(\omega^{'})}
{\omega^{'2}-\omega^{2}}{d\omega^{'}}
\end{equation}
where,\ P stands for\ the principal\ value\ 
of\ the integral.\ The optical\ absorption\ 
coefficient\ was\ obtained by\ using\ 
Eq.~(\ref{eq4})
\begin{equation}{\label{eq4}}
\alpha=\sqrt{2}\frac{\omega}{c}
\sqrt{\sqrt{\varepsilon\rm^{2}_{1}
(\omega)+\varepsilon\rm^{2}_{2}
(\omega)}-\varepsilon\rm_{1}(\omega)}
\end{equation}
where\ $\omega$ is\ photon\ frequency,\ 
and $c$\ is speed\ of light.\ 
$\varepsilon\rm_{1}$ $\&$ $\varepsilon\rm_{2}$\ 
are\ frequency dependent\ real and imaginary\ 
parts of\ dielectric function\ as stated in\ 
Eq.~(\ref{eq4}).\ 
From\ dielectric\ function,\ 
all the other\ optical\ properties\ such as,\ 
reflectivity\ $R$,\ refractive\ index\ $n$,\ 
$\&$\ extinction\ coefficient\ $\kappa$\ is\ 
also\ obtained~\cite{AND2021,SHDYLW2015,TJCMZ2018}.\ 
Refractive index is calculated by 
\begin{equation}{\label{eq5}}
n(\omega)=\frac{1}{\sqrt{2}}
\sqrt{\sqrt{\varepsilon\rm_{1}^{2}
(\omega)+\varepsilon\rm_{2}^{2}
(\omega)}+\varepsilon\rm_{1}(\omega)}
\end{equation}
while\ reflectivity\ $R(\omega)$,\ $\&$\ 
energy\ loss\ function\ $\iota(\omega)$\ 
is\ calculated\ as\ 
\begin{equation}{\label{eq6}}
R(\omega)=\left|\frac{\sqrt{\varepsilon(\omega)}-1}
{\sqrt{\varepsilon(\omega)}+1}\right|^{2}
\end{equation}
and
\begin{equation}{\label{eq7}}
\iota(\omega)=\frac{\varepsilon\rm_{2}(\omega)}
{{\Bigg[}\varepsilon\rm_{1}^{2}(\omega)+
\varepsilon\rm_{2}^{2}(\omega){\Bigg]}},
\end{equation}
respectively.\ 

\section{Results and Discussion \label{sec:res}}
Since ZnO exhibits various structures,\ 
making duly comparison\ between structures on\ 
the relative stabilities\ and the conditions\ 
is necessary\ while studies of\ heterojunction\ 
design\ is considered.\
Some of the bulk properties of ZnO crystal 
is studies by calculating Cohesive energies, 
formation energies, and bulk modulus,\ 
described as follows.\ 
The cohesive and formation energies are 
calculated according to formula given 
our previous work~\cite{SK2022}.\ 
Accordingly,\ as shown\ 
in Table~\ref{tab1},\ the cohesive energy\ of bulk ZnO\ 
structure in WZ\ (B$\rm_{4}$)\ $>$\ ZB\ (B$\rm_{3}$)\ $>$\ 
RS\ (B$\rm_{1}$).\ 
That means\ ZnO is more stable\ in wurtzite~(WZ)\ 
structure,\ followed by\ zincblende (BZ) structure\ 
and is least stable\ for rocksalt (RS) structure.\ 
From energy\ per atom calculation,\
WZ (B$\rm_{4}$)\ $>$\ ZB\ (B$\rm_{3}$)\ $>$\ 
RS\ (B$\rm_{1}$).\ Furthermore,\ the degree of\ 
covalency\ is shown from\ charge values to be\ 
according to WZ (B$\rm_{4}$)\ $>$\ ZB\ (B$\rm_{3}$)\ $>$\ 
RS\ (B$\rm_{1}$).\ 
Positive values\ of $E\rm_{coh}$ in\ 
means exothermic process\ 
while negative values\ mean endothermic\ 
process.\ 
\begin{table}[!ht]
\centering
\addtolength{\tabcolsep}{3.2mm}
\renewcommand{\arraystretch}{1.3}
\caption{Cohesive energy~[eV/atom],\ formation energy~[eV],\ 
         band gap~[eV],\ bulk modulus~[GPa] and lattice\ 
         constant~[{\AA}]\ of bulk ZnO.}
\label{tab1}
\begin{tabular}{|l|l|c|c|c|}
\hline
Quantity & Source & \multicolumn{3}{c|}{Structure}\\
\hline
& & WZ (B$\rm_{4}$) & ZB (B$\rm_{3}$) & RS (B$\rm_{1}$)\\ 
\cline{3-5}
\multirow{2}*{E$\rm_{coh}$} & This work & 2.84 & 1.10 & 0.94\\ 
\cline{2-5}
                       & Expt~value & 1.93~\cite{WRH97} & 0.96~\cite{WRH97} & 0.93~\cite{WRH97}\\ 
\hline
\multirow{2}*{E$\rm_{f}$} & This work & -1.25 & 2.70 & 2.86\\
\cline{2-5}
                     & Expt~value & -3.70~\cite{WRH97} & - & -\\ 
\hline
\multirow{2}*{Bulk modulus} & This work & 137.6 & 131.3 & 169.9\\
\cline{2-5}
                            & Expt~value & 142.4~\cite{DHP98} & - & -\\
\hline
\multirow{2}*{Band gap} & This work & 3.23 & 3.18 & 3.99\\ 
\cline{2-5}
                        & Expt~value &  &  & - \\
\hline
\multirow{2}*{Lattice const~(a)} & This work & 3.4 & 4.4 & 4.1\\ 
\cline{2-5}
                            & Expt~value & 3.25 & - &-\\
\hline 
\multirow{2}*{Lattice const~(c)} & This work & 5.14 & - & -\\ 
\cline{2-5}
                            & Expt~value & 5.2 & - & -\\
\hline
\multirow{2}*{Energy per atom} & This work & -3.32 & -1.58 & -1.41\\ 
\cline{2-5}
                            & Expt~value & - & - & -\\
\hline
\multirow{2}*{{$\Delta$}Q[$e$] on Zn} & This work & +0.23 & +0.25 & +0.41\\ 
\cline{2-5}
                            & Expt~value & - & - & -\\
\hline
\multirow{2}*{{$\Delta$}Q[$e$] on O} & This work & -0.23 & -0.25 & -0.41\\ 
\cline{2-5}
                            & Expt~value & - & - & -\\
\hline
\end{tabular}
\end{table}

In terms of\ 
formation energy\ in reference\ to Table~\ref{tab1},\ 
thus,\ $E\rm_{f}$ of bulk ZnO\ structure in\ 
WZ~(B$\rm_{4}$)~$<$~ZB~(B$\rm_{3}$)~$<$~RS~(B$\rm_{1}$).\ 
This\ means ZnO alloy\ is more favorable\ to be\ 
formed\ in wurtzite~(B$\rm_{4}$) structure.\ 
Furthermore,\ based on the\ bulk modulus values\ 
of ZnO\ structures\ presented in Table~\ref{tab1},\ 
WZ~(B$\rm_{4}$)~$>$~ZB~(B$\rm_{3}$)~$>$~RS~(B$\rm_{1}$).\ 
This means,\ ZnO alloy is more\ resistant\ 
to extreme pressure\ conditions in\ wurtzite\ 
structure.\
Negative values\ of $E\rm_{f}$ in\ 
means exothermic process\ 
while positive values\ mean endothermic\ 
process.\  
From a curve fits according to Murnaghan~\cite{FM1944}\ 
approach,\ to an energy\ versus volume\ 
calculation\ data,\ the bulk modulus,\ 
minimum\ volume and minimum energy\ 
values are obtained, as shown in the 
Table~\ref{tab1}.\   

Accordingly,\ the output parameters\ of lattice\ 
constants (a),\ (c),\ and bulk modulus~(B) for\ 
B$\rm_{4}$ structure in this work are\ 3.4~{\AA},\ 
5.14~{\AA},\ and 137.59 GPa\ respectively.\ But\ 
the\ experimental results\ of lattice constants (a),\ (c),\ 
and bulk modulus~(B)\ are 3.25~{\AA}~\cite{RMA01},\ 
5.2~{\AA}~\cite{RMA01},\ and 142.4 GPa~\cite{DHP98},\ 
respectively.\ 
The results\ show\ that even though\ there is small\ 
differences in\ between the\ computational and the\ 
experimental\ results,\ the two are\ essentially\ in\ 
agreement\ with each\ other.\ This gives\ us\
a confidence on the predicted properties\ of\ 
heterojunction structures\ considered in this\ 
study.\  
The surface\ energy ${\sigma}$~[eV/{\AA}$\rm^{2}$]\ 
is defined as\
\begin{equation}
\sigma = \frac{1}{2A}{\Bigg(}E\rm_{slab} 
- \frac{N\rm_{slab}}{N\rm_{bulk}}E\rm_{bulk}{\Bigg)}
\label{eq8}
\end{equation}
where\ $E\rm_{slab}$\ denotes the\ total energy\ 
of a slab\ unit cell,\ $N\rm_{slab}$,\ means number\ 
of atoms in the\ slab unit cell,\ $N\rm_{bulk}$\ 
means number\ of atoms in the\ bulk unit cell,\ 
and $E\rm_{bulk}$\ is the total\ energy of a bulk\ 
unit cell,\ A is the surface\ area of a slab\ 
unit cell.\ As such,\ a high surface\ energy would\ 
indicate increased\ reactivity with\ adsorbates,\
while low\ surface energy\ would indicate\ 
increased stability.\
Our calculations\ using Eq.~\eqref{eq8}\ show that\ 
surface energies\ for the surface\ facets increase\ 
according to\ $(001)<(100)<(110)<(111)$\ for all\ 
the structures.\ This means (001)\ surface offer\ 
a relatively more\ stable geometry,\ while (111)\ 
facet\ would likely\ be expected to be\ more\ 
reactive to impurities / adsorbates.\ 
For the\ clean WZ ZnO~surfaces,\ we obtain\ 
a work function\ value of between 4.3 to 6.8 eV,\ 
while\ the corresponding literature values\ for\ 
ZnO surfaces\ vary\ between 3.7 and 6.0 eV~\cite{JZG84}.\
Among the surface facets considered, we calculated
work function values of 6.8 eV for $(001)$,\ 
4.5 eV for $(100)$,\ 4.3 eV for $(110)$,\
and 4.9 eV for $(111)$.\ In view of this,\ 
we considered\ the $(001)$\ surface as a better\
suit to build-up\ a heterostructure of our\ 
study.\        
\subsection{Electronic and optical properties of clean ZnO}

\begin{figure}[!ht]
\centering
\includegraphics[scale=0.25]{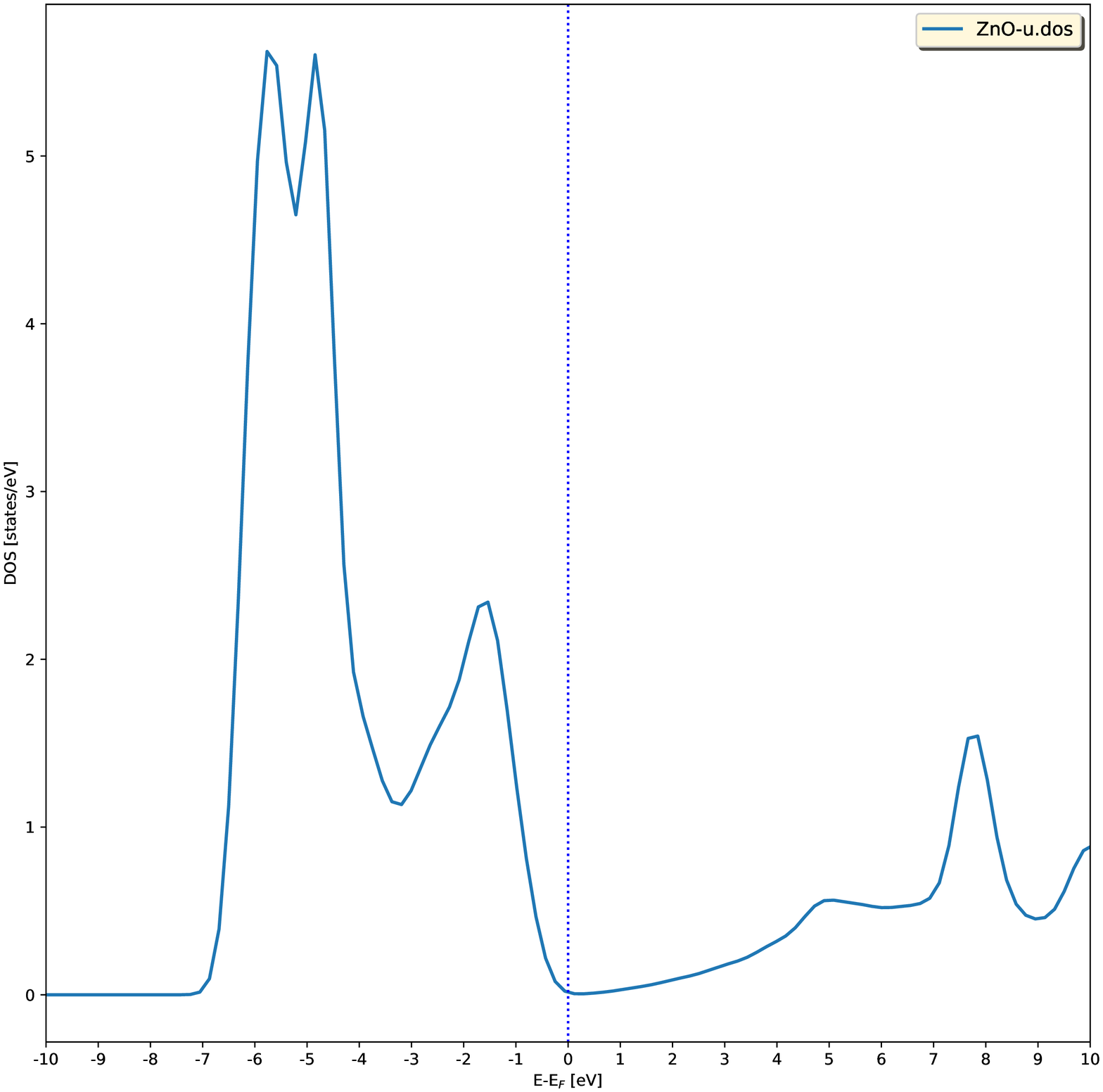}
\includegraphics[scale=0.25]{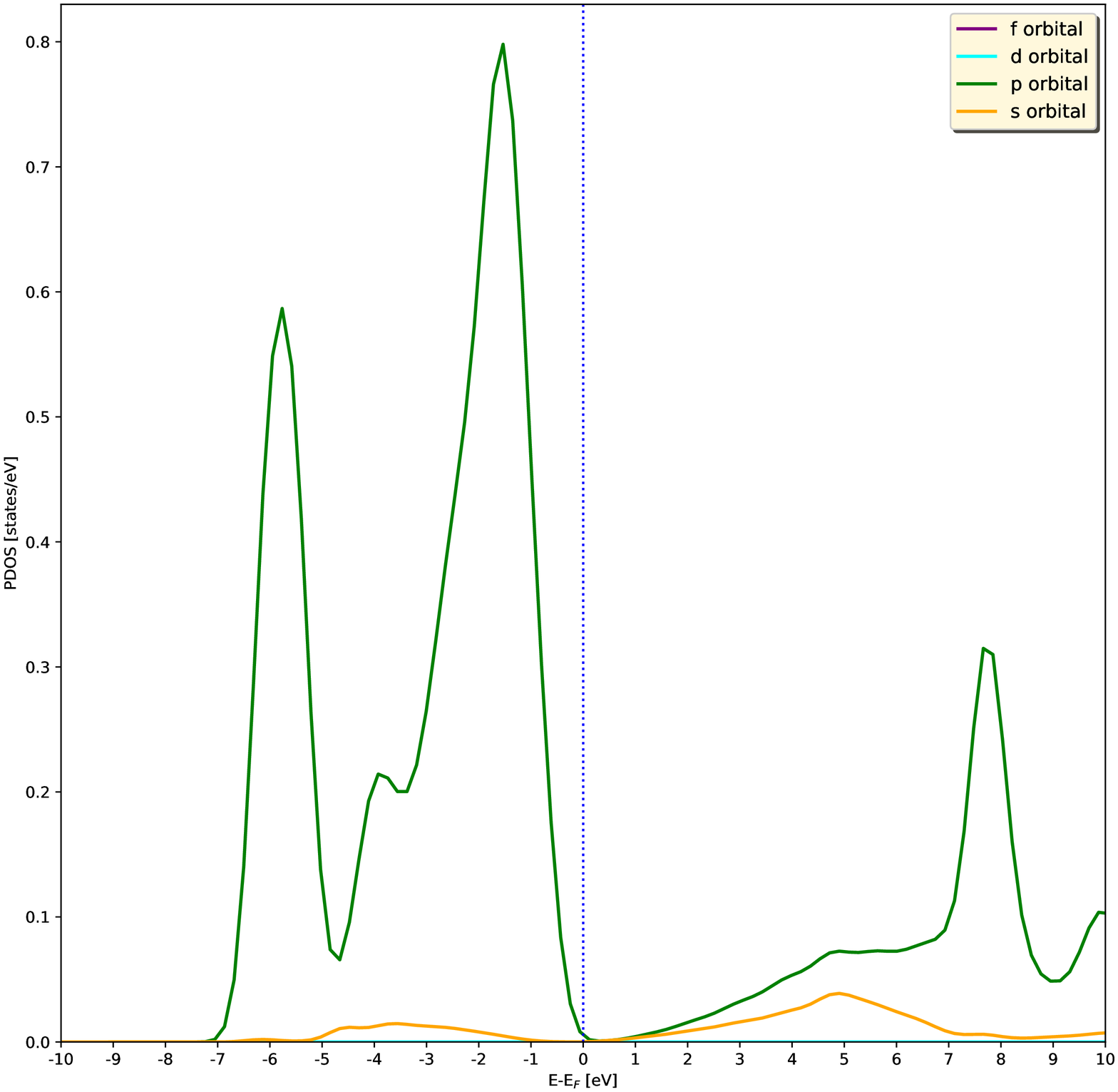}
\caption{DOS and PDOS of bulk wurtzite ZnO.}
\label{fig1}
\end{figure}

As shown in Fig.~\ref{fig1},\ the lower part\ 
of the valence band\ at an energy\ of\ 
${\sim}$~-7~eV to ${\sim}$~-4~eV shows narrow\ 
sharp peaks in\ DOS. 
The projected density of states~(PDOS)\ is the\ 
relative contribution of a particular orbital to\ 
the DOS.\ As shown in Fig.~\ref{fig1},\ the $p$ and $s$\ 
orbitals contribute\ most to the DOS curve.\
The $p$ orbital has\ most states for occupation\ 
followed\ by $s$ orbital.\ The $p$ states dominate\ 
near the Fermi level,\ while the $s$ states contributes\ 
in the deepest\ energy level.\

\begin{figure}[!ht]
\centering
\includegraphics[scale=0.45]{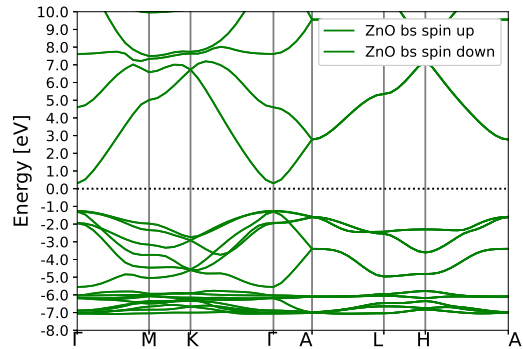}
\caption{Band structure of bulk wurtzite ZnO.}
\label{fig2}
\end{figure}

From Fig.~\ref{fig2},\ it can be seen\ that the\ 
valence band\ maxima and conduction\ band minima\ 
show smaller\ bandgaps at\ identical k-points,\ 
indicating\ direct band.\ The band structure\ 
in the figure\ for wurtzite bulk ZnO,\ 
point ${\Gamma}-{\Gamma}$\ showed the\ lowest\ 
energy band\ gap,\ which is\ $\sim$~3.23~eV.\ 
The excitation\ of electrons\ 
(from the highest valence\ band to the lowest\ 
conduction band)\ will occur\ at the lowest\ 
energy band gap.\ Thus,\ point (${\Gamma}-{\Gamma}$)\ 
is the point\ where electron\ excitation most\ 
probably occurred.\ The computed band gap\ 
is in agreement\ with the\ experimental\ 
value 3.37~eV.\ 


\begin{figure}[!ht]
\centering
\includegraphics[scale=0.5]{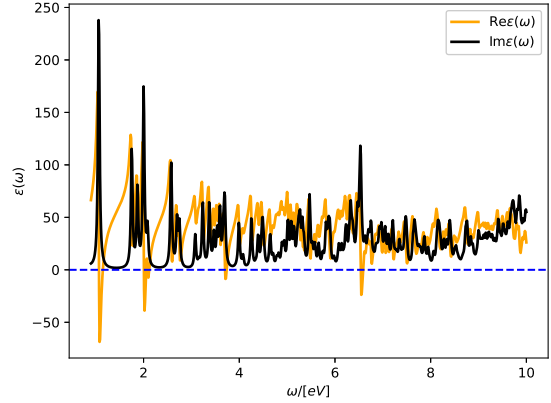}
\caption{The real~(yellow color line) and imaginary\ 
         (black color line) parts of frequency-dependent\ 
         complex dielectric function of wurtzite bulk ZnO\ 
         with local field effects.}
\label{fig3}
\end{figure} 
As shown in Fig.~\ref{fig3},\ the imaginary part\ 
${\varepsilon_{2}}~(\omega)$~(absorptive part)\ of the\ 
dielectric function (black color line)\ 
illustrates the optical\ transition mechanism.\ Each peak\ 
in the imaginary\ part of the dielectric\ function\ 
corresponds to\ an electronic transition.\ The highest\ 
peak of\ $\varepsilon_{2}(\omega)$\ is\ located\ 
at\ 0.61~eV,\ with the value\ of $\varepsilon_{2}(\omega)$\ 
is 36.69.\ At zero\ photon energy,\ $\varepsilon_{2}(0)$\
has a value of 10.5.\

\begin{figure}[!ht]
\centering
\includegraphics[scale=0.35]{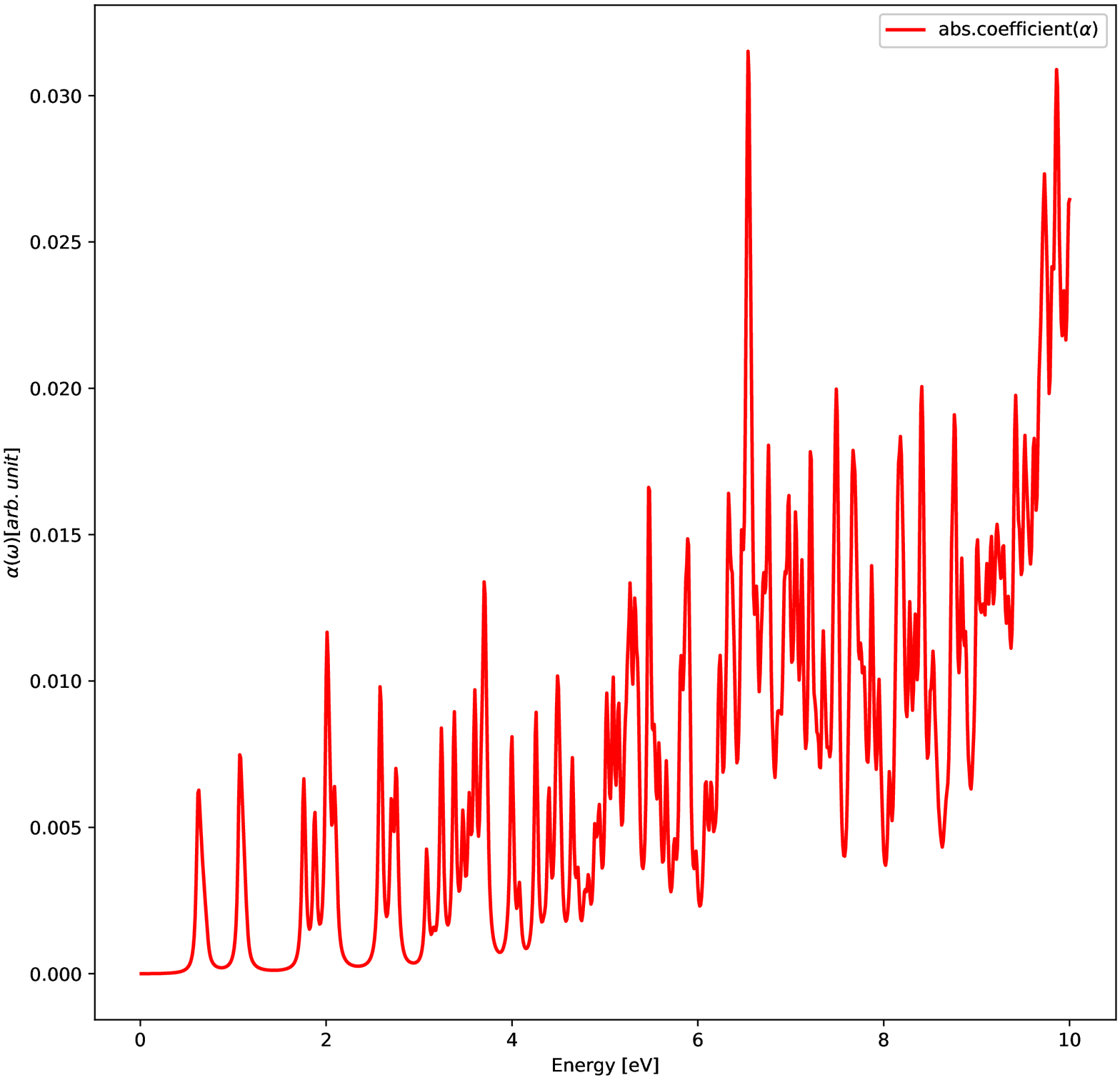}\\
\includegraphics[scale=0.35,angle=270]{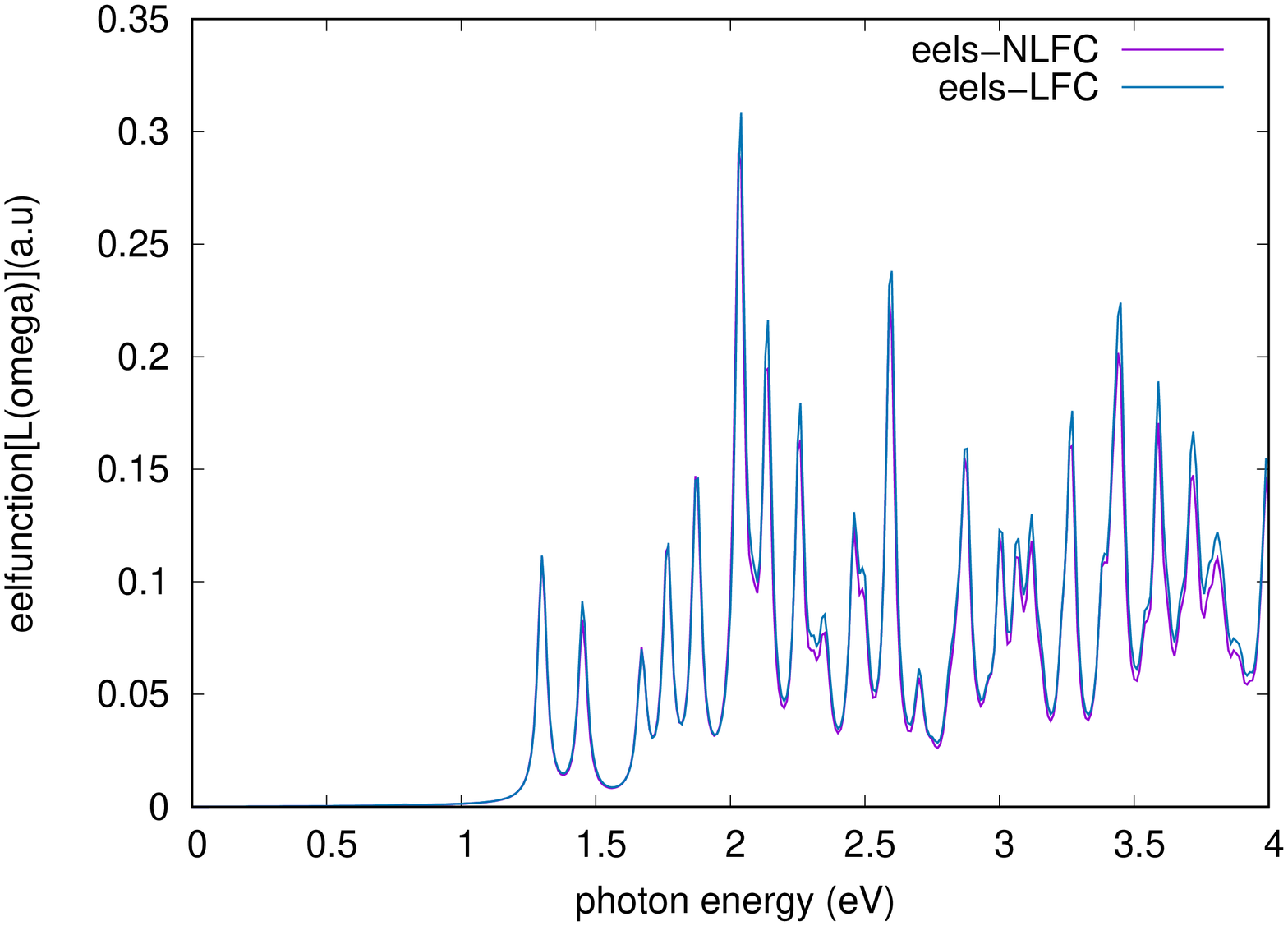}
\caption{Absorption coefficient and electron 
         energy loss of bulk wurtzite\ 
         ZnO as a function of energy.}
\label{fig4}
\end{figure} 
The absorption\ coefficient (Fig.~\ref{fig4})\
shows peaks at\ ranges of\ photon energies\
including at 0.6 eV, 1.2 eV, 1.8 eV, 2.1 eV,\ 
2.5 eV,\ and 3.5 eV.\  
These correspond\ to absorptions of\ near-infrared,\
visible,\ and near-ultraviolet lights.\ 
Meanwhile,\ the electron\ energy loss\ 
curve\ shows\ occurrences at\ photon energies\ 
exceeding 1.0 eV.\   


\begin{figure}[!ht]
\centering
\includegraphics[scale=0.20]{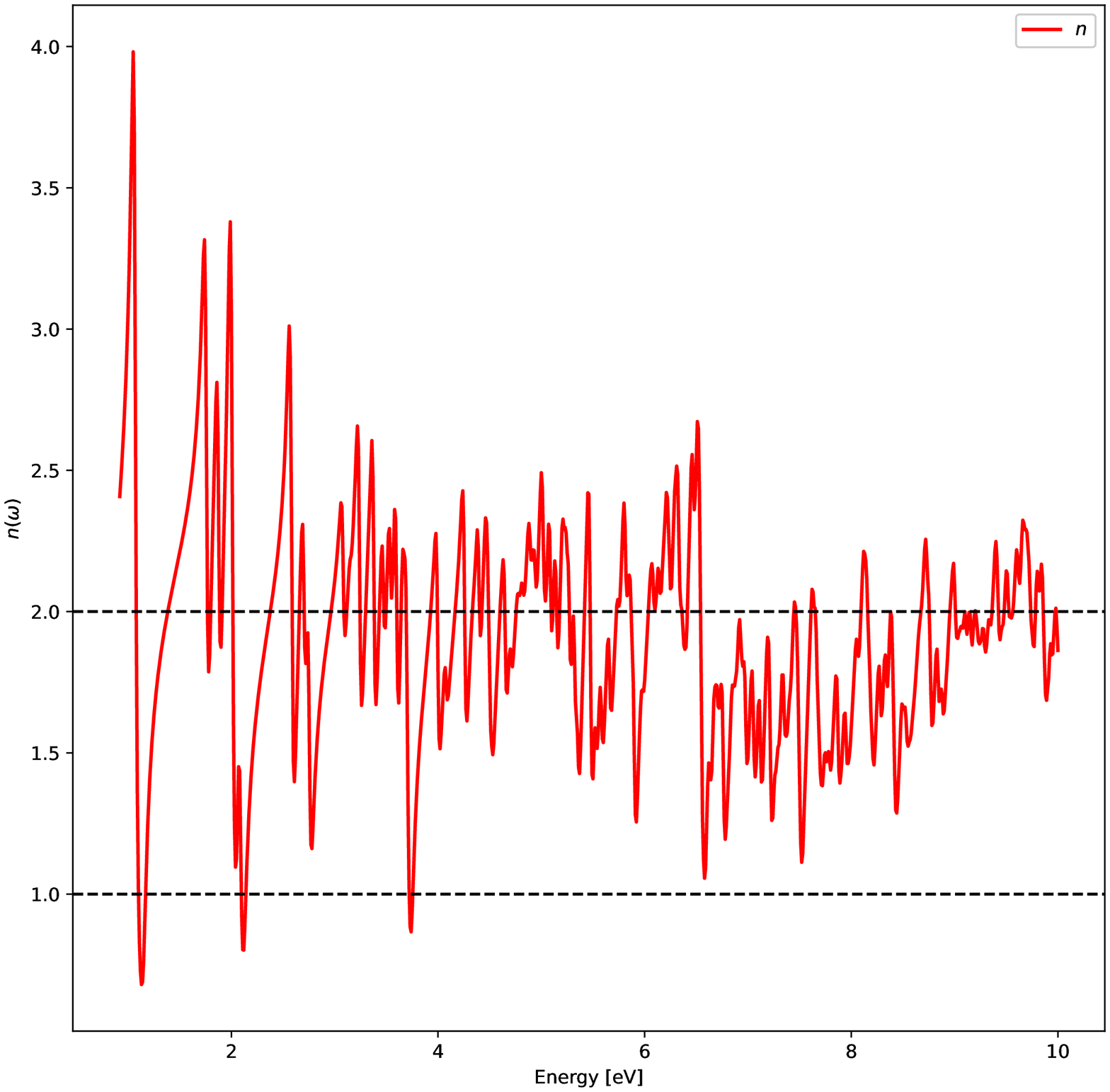}
\includegraphics[scale=0.45]{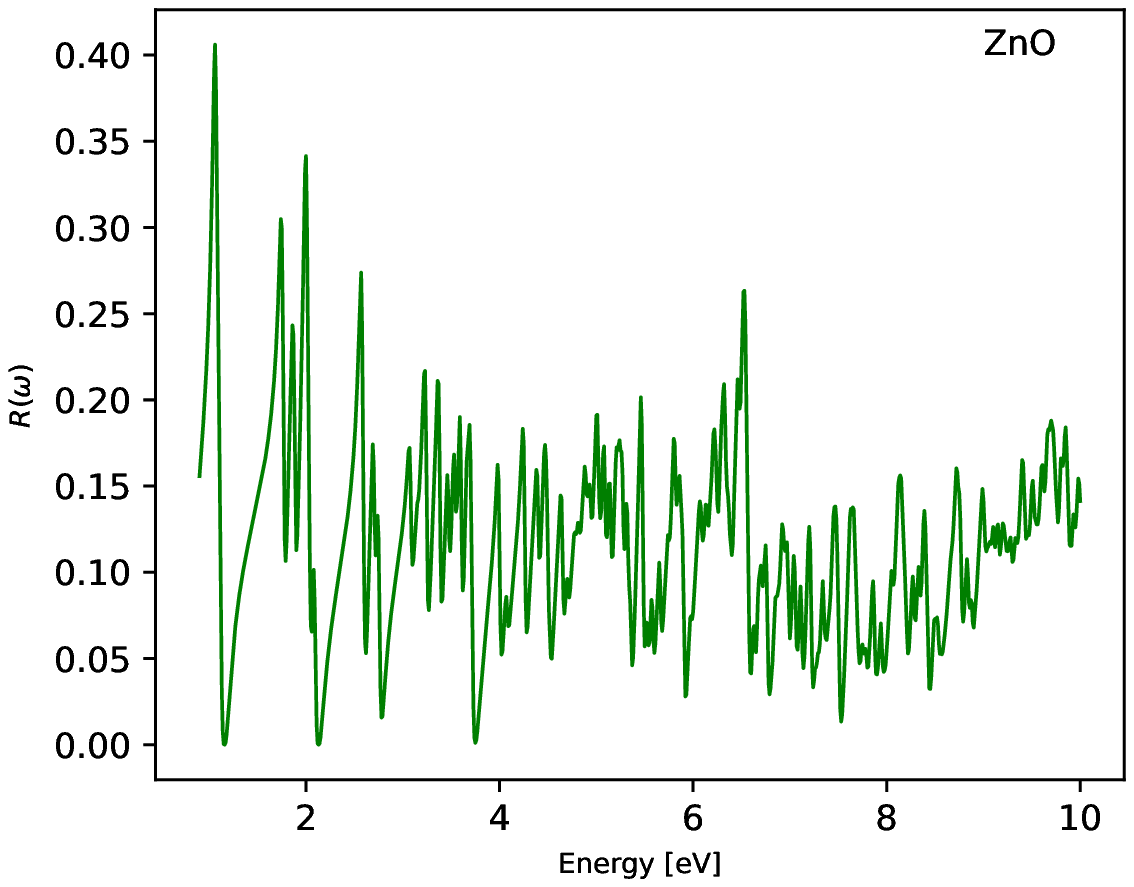}
\caption{Index of refraction and reflectivity 
         of bulk wurtzite\ 
         ZnO as a function of energy.}
\label{fig5}
\end{figure} 

Figure~\ref{fig5}\ shows\ the refraction 
and reflectivity properties of the wurtzite\ 
ZnO.\ The propagation of\ a light beam through a\ 
translucent medium via refraction is\ described\ 
by the\ refractive index $n$.\ The static\ values\ 
of the index\ of\ refraction,\ $n(0)$,\ is 2.4.\
The reflectivity\ represents\ propagation\
via\ reflection.\ The static\ values of the reflection\ 
coefficient,\ $R(0)$,\ is 0.4.\
\subsection{Structural and electronic properties of graphene/ZnO heterojunction}
\subsubsection{Structural properties of graphene/ZnO hetrojunction}
\begin{figure}[!hbtp]
\centering
\includegraphics[scale=0.5]{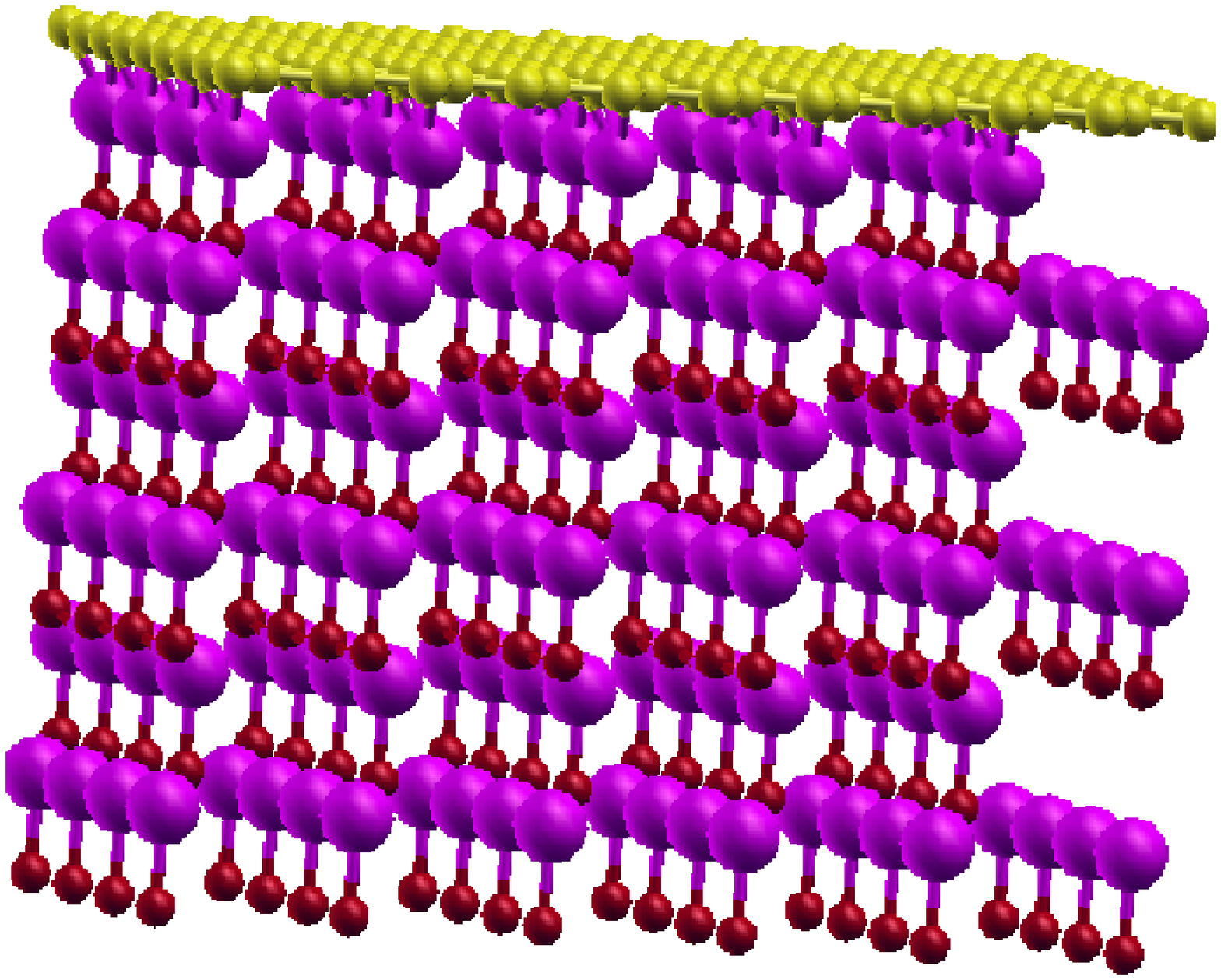}
\caption{Graphene/ZnO heterojunction structure. 
         Color Online. Colors: C-yellow, Zn-pink,\ O-red.}
\label{fig6}
\end{figure}

We considered\ the adsorption of\ graphene on ZnO\ 
polar surfaces.\ The adsorption energy\ of graphene\ 
layer\ is defined\ by\ 
\begin{equation}
E_{ads}=E_{interface}-E_{graphene}-E_{ZnO}
\label{eq9}
\end{equation}
while,\
\begin{equation}
e_{ads}=\frac{{E_{interface}}-{E_{graphene}}-{E_{ZnO}}}{n}
\label{eq10}
\end{equation}
where $e\rm_{ads}$~[eV/atom] is\ the adsorption energy\ 
per C\ atom;\ $E\rm_{interface}$,\ $E\rm_{graphene}$,\ and\ 
$E\rm_{ZnO}$ is the total\ energies of the\ interface,\ 
graphene,\ and ZnO surface,\ respectively;\ $n$ is\ 
the total number\ of carbon\ atoms in\ the interface.\ 
Figure~\ref{fig6} shows\ that the adsorption\ 
behaviors of\ the three\ configurations for\ 
ZnO(0001)\ and $(000\bar{1})$\ surfaces are\ 
almost the same.\

More importantly,\ the long range\ vdW interaction\ 
plays an important\ role in the adsorption.\ Both\ 
physical and chemical\ adsorptions\ take place\ 
when graphene\ adheres to\ bare SiC~\cite{JTM2011},\ 
SiO$\rm_{2}~$\cite{FWV2012},\ and metals~\cite{GLS2010}\ 
surfaces.\ However,\ no chemical adsorption\ is\ 
seen when\ graphene adheres\ to ZnO surfaces.\ 
For use in\ photocatalysts\ and solar cells,\ 
the intimate\ but nondestructive\ contact between\ 
graphene and ZnO\ may be unique.\
The shortest separation\ distance between an\ 
atom of the adsorbed\ ZnO and\ the closest C\ 
atom of the graphene\ monolayer in the\ 
graphene/ZnO is\ 2.52~{\AA},\ which is in\ 
agreement with\ a literature result~\cite{BGK11}.\
It looks that\ some sort of charge\ transfer from\ 
graphene layer\ to the underlying ZnO surface\ 
takes place\ at such optimum\ separation distance,\ 
while also\ the corresponding energy\ per atom\ 
and cohesive energies\ show stabilities\ of the\ 
structure.
\begin{figure}[!hbtp]
\centering
\includegraphics[scale=0.35]{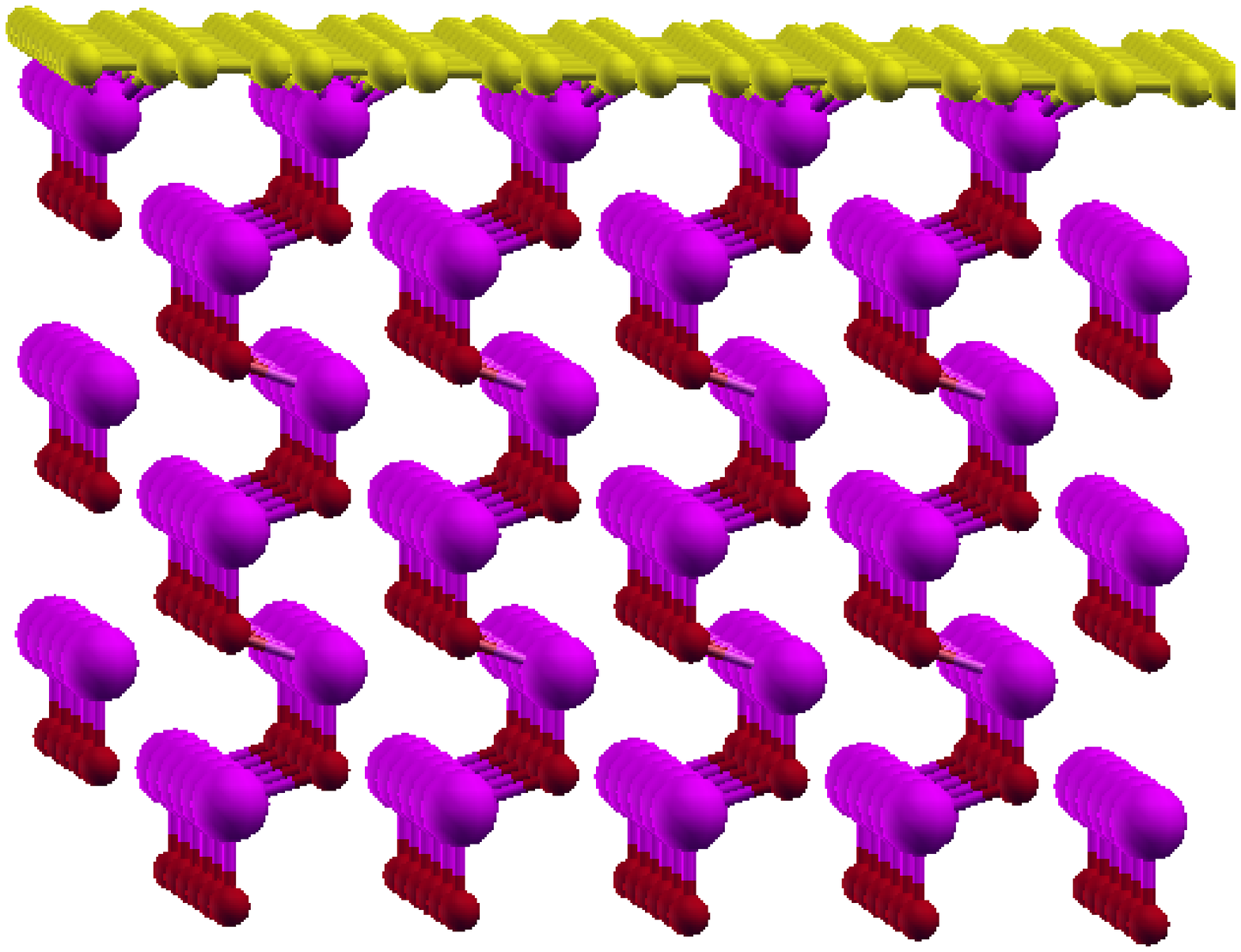}
\includegraphics[scale=0.35]{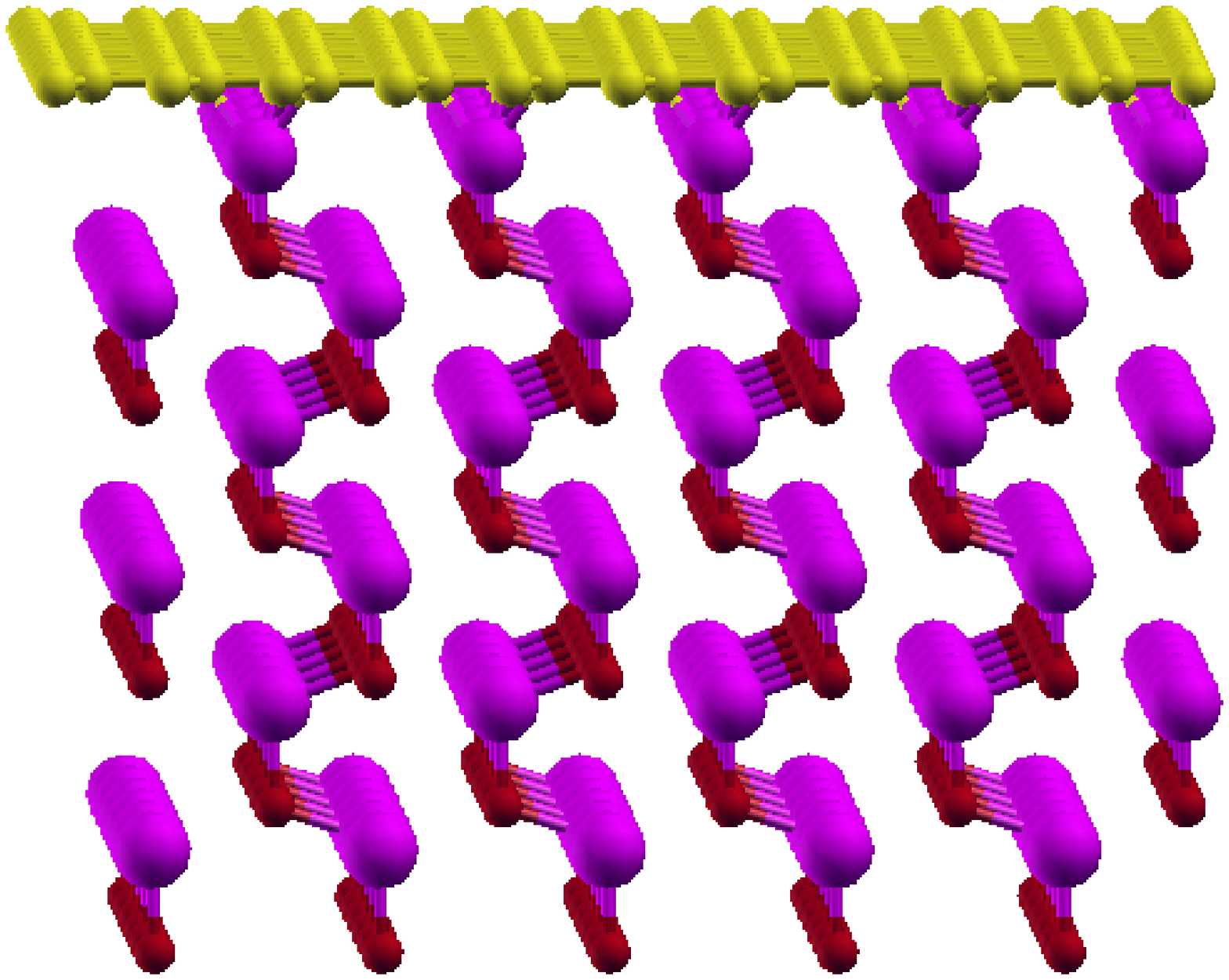}
\includegraphics[scale=0.5]{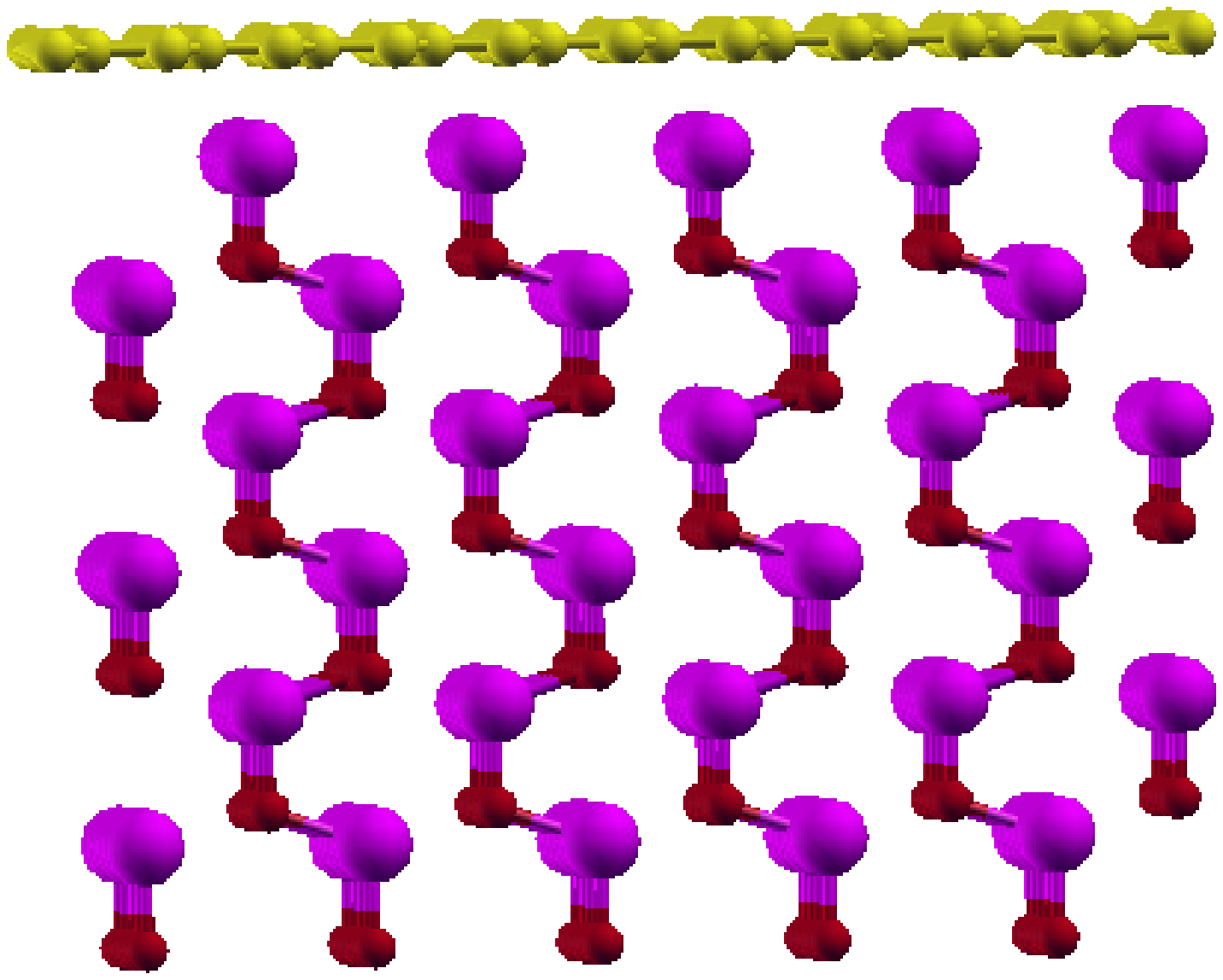}
\includegraphics[scale=0.4]{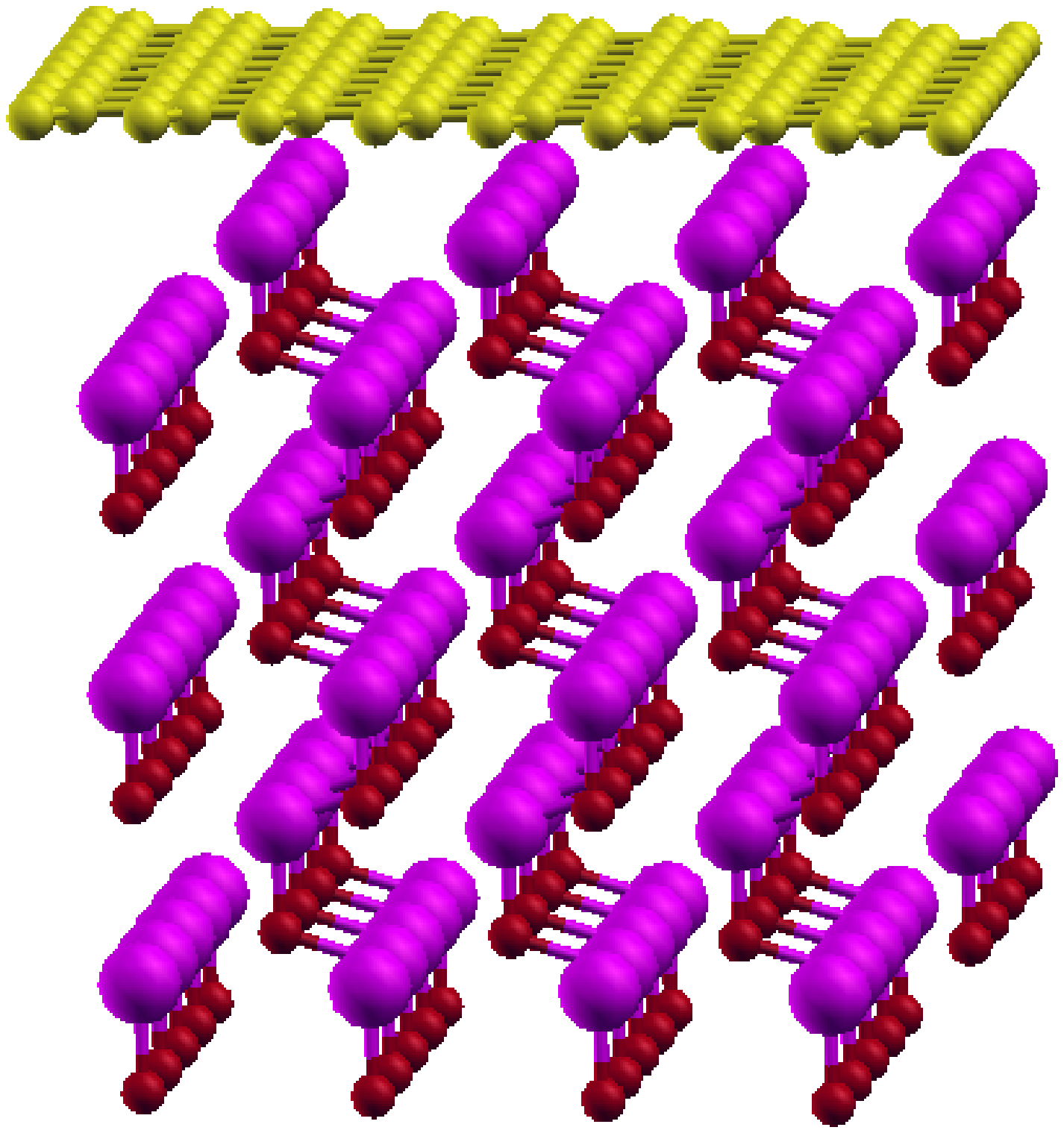}
\caption{ZnO-001-clean-T-graphene at different equilibrium distance.
         Top to bottom, $d=1.02$~{\AA},\ $d=1.52$~{\AA},\ 
         $d=2.12$~{\AA},\ $d=2.52$~{\AA},\ respectively.\
         Color Online. Colors: C-yellow, Zn-pink,\ O-red.}
\label{fig7}
\end{figure}
\begin{table}[!hbtp]
\centering
\addtolength{\tabcolsep}{0.4mm}
\renewcommand{\arraystretch}{1.3}
\caption{Adsorption energy per atom,\ cohesive energy,\ 
         and $\Delta$Q (charge transfer from graphene
         to the interface)\ as a function\ 
         of $\frac{{d_{0}}-d}{d_{0}}$.}
\label{tab2}
\begin{tabular}{|l|c|c|c|c|c|c|c|c|}
\hline
Quantity & \multicolumn{8}{c|}{$\frac{{d_{0}}-d}{d_{0}}$}\\
\cline{2-9}
& 0.99 & 0.79 & 0.59 & 0.39 & 0.27 & 0.00 & 0.15 & -0.19\\
\hline
Energy per atom~[eV] & -18.31 & -14.19 & -9.92 & -8.60 & -8.43 & -8.39 & -8.38 & -8.36\\
\hline
Cohesive Energy~[eV] & 1.25 & 2.9 & 4.61 & 5.13 & 5.21 & 5.22 & 5.22 & 5.23\\
\hline
$\Delta$Q~[$e$] & +0.63 & +0.19 & -0.38 & -0.33 & -0.23 & -0.13 & +0.05 & +0.11\\
\hline
\end{tabular}
\end{table}
\begin{figure}[!ht]
\centering
\includegraphics[scale=0.3]{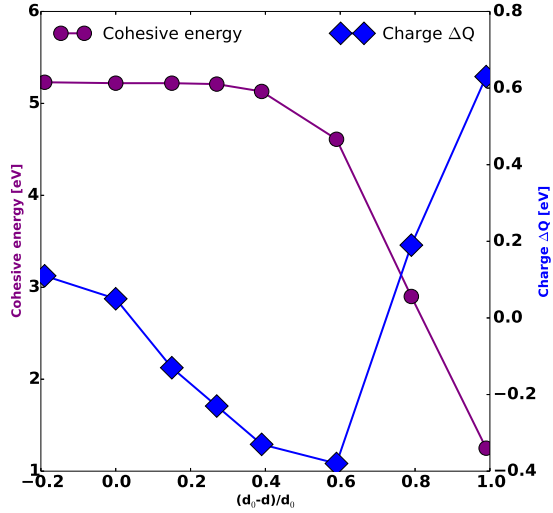}
\caption{Cohesive energy and charge transfer $\Delta$Q\ 
         as a function of $\frac{{d_{0}}-d}{d_{0}}$.\
         Color Online.\ Purple color:~Cohesive 
         energy, and blue color:~$\Delta$Q.}
\label{fig8}
\end{figure}
\begin{figure}[!ht]
\centering
\includegraphics[scale=0.3]{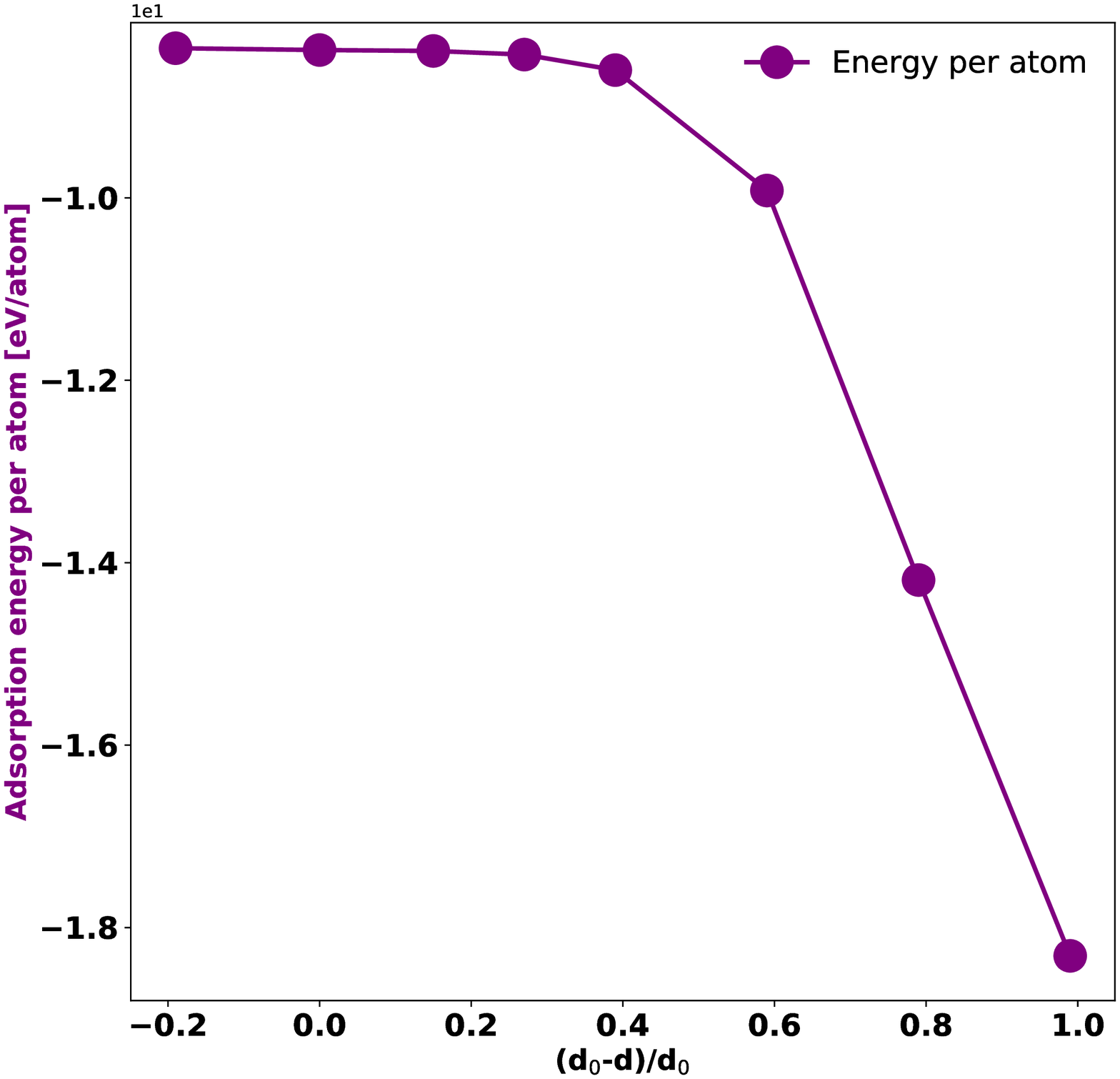}
\caption{Adsorption energy\ per atom as\ a function\ 
         of $\frac{{d_{0}}-d}{d_{0}}$.}
\label{fig9}
\end{figure}
The geometries\ of interface\ structures of\ 
the heterojunction\ is given in Fig.~\ref{fig7},\ 
while\ the corresponding\ graphic plots\ of the\ 
cohesive\ and adsorption\ energies trends\ are\ 
given in\ Figs.~\ref{fig8}~$\&$~\ref{fig9},\ 
respectively.\ Table~\ref{tab2}\ gives\ quantitative\ 
values\ of\ the adsorption and\ cohesive energies.\  
\subsubsection{Electronic properties of graphene/ZnO heterojunction}
\begin{figure}[!hbtp]
\centering
\includegraphics[scale=0.3]{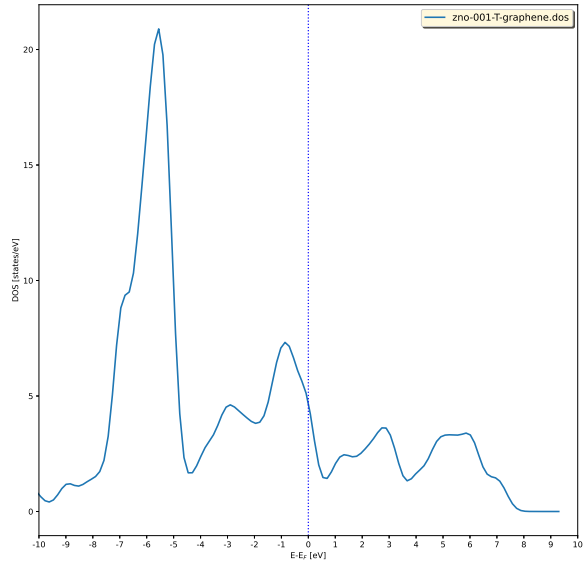}
\caption{DOS of graphene/ZnO heterojunction.}
\label{fig10}
\end{figure}

\begin{figure}[!hbtp]
\centering
\includegraphics[scale=0.3]{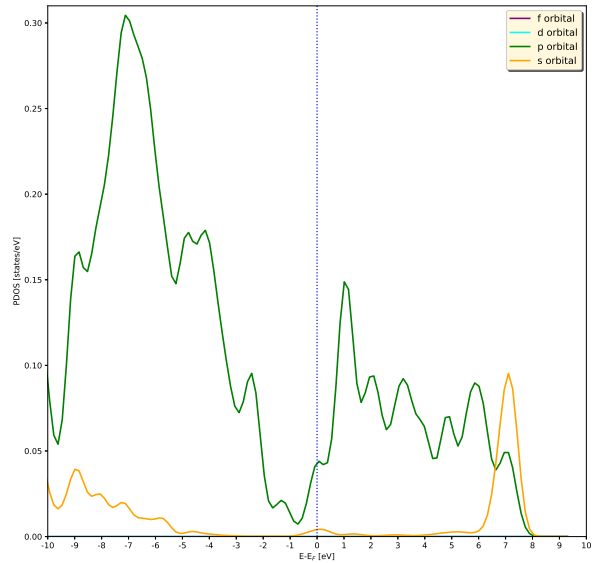}
\caption{PDOS of graphene/ZnO heterojunction.}
\label{fig11}
\end{figure}

As shown in Fig.~\ref{fig10},\ the graphene/ZnO\ 
heterojunction shows\ a narrow sharp\ peak in the\ 
DOS at\ $\sim$-8 eV to $\sim$-4.5 eV\ which is\ 
associated with\ a large quantity\ of generated\ 
states in\ the valence band\ and nonzero DOS at\ 
the Fermi energy,\ indicating that\ graphene/ZnO\ 
has a metallic character.\ 
As PDOS is\ the relative contribution\ of a particular\ 
atom/orbital to\ the DOS,\ as\ shown\ in Fig.~\ref{fig11},\ 
the $p$ and $s$ orbitals\ contribute\ to the occupation\ 
states of DOS.\ The $p$ orbital has\ a major contribution\ 
and the $s$ orbital\ has a minor contribution\ to the DOS\ 
in both the\ conduction and valence bands.\  
\begin{figure}[!ht]
\centering
\includegraphics[scale=0.4]{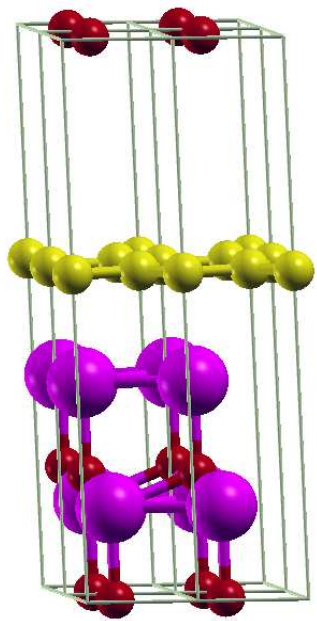}
\caption{3D view of\ heterojunction\ structure.\ 
         Color Online.\ Colors: C-yellow,\ 
         Zn-pink,\ and O-red.}
\label{fig12}
\end{figure}

The charge calculations\ on the atoms of the\ 
unit cell shows that\ 0.05 electrons\ 
have been\ transferred from\ graphene atom to\ 
the\ Zinc Oxide.\ This shows that\ at the junction,\ 
graphene become $p$-type\ and ZnO is\ $n$-type material.\ 
A\ summary of the\ charge transfers\ for different\ 
layer separation\ distances is\ given in Table~\ref{tab2}\ 
$\&$ Fig.~\ref{fig8}.\ It\ seems that the\ 
heterojunction increases\ the electrical property,\
(see Figs.~\ref{fig10}~$\&$~\ref{fig11}),\ while\ 
also likely improving\ its optical properties.\ 
Figure~\ref{fig12}\ shows\ the 3D\ counterpart\ 
of the 2D layer\ interface presentations of\ 
Fig.~\ref{fig7}.\ It looks from\ 
Figs.~\ref{fig8}~$\&$~\ref{fig9}\ that the\ 
cohesive energies\ and adsorption energies\ per atom\
show saturated\ values beginning\ from optimum\ 
separation distance\ of ${d_{0}}=2.52$~{\AA}.\ 

\section{Conclusion\label{sec:conc}}
The calculation\ of the bulk structural,\ 
electronic and optical\ properties of graphene\ 
and ZnO\ reveals\ various\ characteristics.\ The\ 
structure,\ including equilibrium\ lattice\ 
constants,\ bulk modulus,\ cohesive,\ and\ 
formation energies\ are in good agreement\ 
with other experimental results.\ The electronic\ 
properties\ are studied using band gap,\ band structures,\ 
DOS/PDOS,\ and charge analysis.
Surface properties\ of clean\ wurtzite ZnO polar\ 
surfaces were analysed by\ calculating the surface\ 
energy\ and work functions.\\ 

The surface energies\ for the surface\ facets increase\ 
according to $(001)<(100)<(110)<(111)$\ for all\ 
the structures.\ Based on the outcomes,\ we\
suggest that (001) surfaces offer a\ relatively more\ 
stable geometries,\ while (111) facets would\ 
likely be expected\ to be more reactive\ to\ 
impurities/adsorbates.\ As\ a result,\ we have\ 
chosen\ the polar (001)\ surface to be suitable\
for forming\ the\ graphene/ZnO heterostructure.\ 
The electronic properties\ of graphene/ZnO\ 
heterostructure\ is\ revealed\ in\ bader charge,\
DOS,\ and PDOS analysis.\ The results show\ that\
there is a\ charge transfer\ in between graphene\ 
and ZnO and\ the combination/junction\ seems\
to\ show\ zero\ band gap.\
Furthermore,\ our findings seem to offer\ 
compelling justifications for the enhanced\ 
photocatalytic efficiency of graphene/ZnO\ 
hybrid materials.\

\section*{CRediT authorship contribution statement}
H.D.~Etea\ conducted the\ DFT calculations,\
and wrote the draft manuscript,\ and K.N.~Nigussa\ 
carried out the\ research process\ and the\ 
revised writing of the manuscript.\   
\section*{Declaration of Competing Interest}
The authors declare\ that they have no known\ 
competing financial\ interests or personal\ 
relationships that\ could have appeared\ 
to influence the\ work reported in this paper.\

\section*{Acknowledgments}
We are grateful to the Ministry of Education\ 
of\ Ethiopia for financial support.\ The\ 
authors also acknowledge\ the\ Department of\ 
Physics at\ Addis Ababa University.\ 
The~office~of~VPRTT~of Addis\ Ababa\ University\ 
is also warmly~appreciated~for\ supporting~this\ 
research under a\ grant~number~AR/053/2021.\ 

\section*{Data\ Availability\ Statement}
The data that\ support the findings\ of\ 
this study\ are available\ upon reasonable\ 
request\ from the\ authors.\
\bibliographystyle{elsarticle-num}
\bibliography{refs.bib}
\end{document}